\documentclass[english,aip,apl,reprint]{revtex4-1}
\usepackage{units}
\usepackage{color}
\usepackage{amstext}
\usepackage{amssymb}
\usepackage{amsmath}
\usepackage{graphicx}

\begin{document}

\title{Pauli spin blockade in undoped Si/SiGe two-electron double quantum dots}

\author{M.~G.~Borselli}
\email{mborselli@hrl.com}
\author{K.~Eng}
\author{E.~T.~Croke}
\author{B.~M.~Maune}
\author{B.~Huang}
\author{R.~S.~Ross}
\author{A.~A.~Kiselev}
\author{P.~W.~Deelman} 
\author{I.~Alvarado-Rodriguez}  
\author{A.~E.~Schmitz} 
\author{M.~Sokolich} 
\author{K.~S.~Holabird} 
\author{T.~M.~Hazard}  
\author{M.~F.~Gyure}
\author{A.~T.~Hunter}
\noaffiliation

\affiliation{HRL Laboratories,\,LLC, Malibu, CA 90265, USA}

%\date{\today}
\date{30 June 2011}

\begin{abstract}
We demonstrate double quantum dots fabricated in undoped Si/SiGe heterostructures relying on a double top-gated design. Charge sensing shows that we can reliably deplete these devices to zero charge occupancy. Measurements and simulations confirm that the energetics are determined by the gate-induced electrostatic potentials. Pauli spin blockade has been observed via transport through the double dot in the two electron configuration, a critical step in performing coherent spin manipulations in Si.
\end{abstract}

\maketitle

Quantum dots in silicon are currently the focus of many research efforts due to their potential use for semiconductor-based quantum information processing.\cite{Eriksson2004} Measurements of several important properties of Si quantum dots have been recently reported such as spin relaxation,\cite{Morello2010,Hayes2009,Xiao2010PRL,Simmons2011} valley splitting,\cite{Borselli2011,Xiao2010APL,Lim2011} and spin blockade.\cite{Liu2008,Shaji2008,Lai2011} However, existing Si quantum dot designs, based on Si-MOS technology or strained Si/SiGe modulation doped heterostructures, continue to be plagued by the presence of interface charge and ionized dopants. The large effective electron mass in Si, $0.19~m_e$, nearly three times that in GaAs, serves to increase the relative influence of the charge induced disorder potential, requiring the fabrication of exceedingly small devices. An attractive approach is to isolate electrons from these sources of disorder by using an undoped Si/SiGe heterostructure together with a global accumulation gate.\cite{Jackson1993} Devices fabricated on such structures have recently demonstrated high electron mobilities\cite{Lu2009} and Coulomb blockade.\cite{Lu2011}

\begin{figure}
\centering
\includegraphics[width=3.2in]{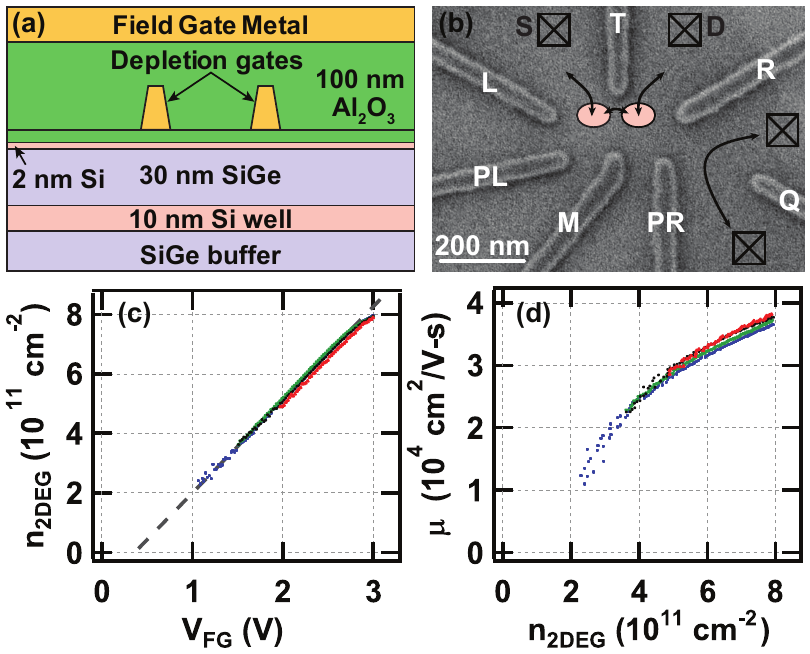}
\caption{\label{fig:Fig1} (a) Schematic cross-section of the device. (b) Scanning electron micrograph of device C after depletion gate metallization. Ohmic connections to 2DEGs are represented by black boxes. Current through the dot is measured between S and D, and QPC current is measured between the rightmost ohmics. Measurements from four Hall bars represented by different symbols (color online) show (c) electron concentration, $n_\text{2DEG}$, versus field gate bias, $V_\text{FG}$, and (d) mobility, $\mu$, vs. $n_\text{2DEG}$.}
\end{figure}

\begin{figure*}
\centering
\includegraphics{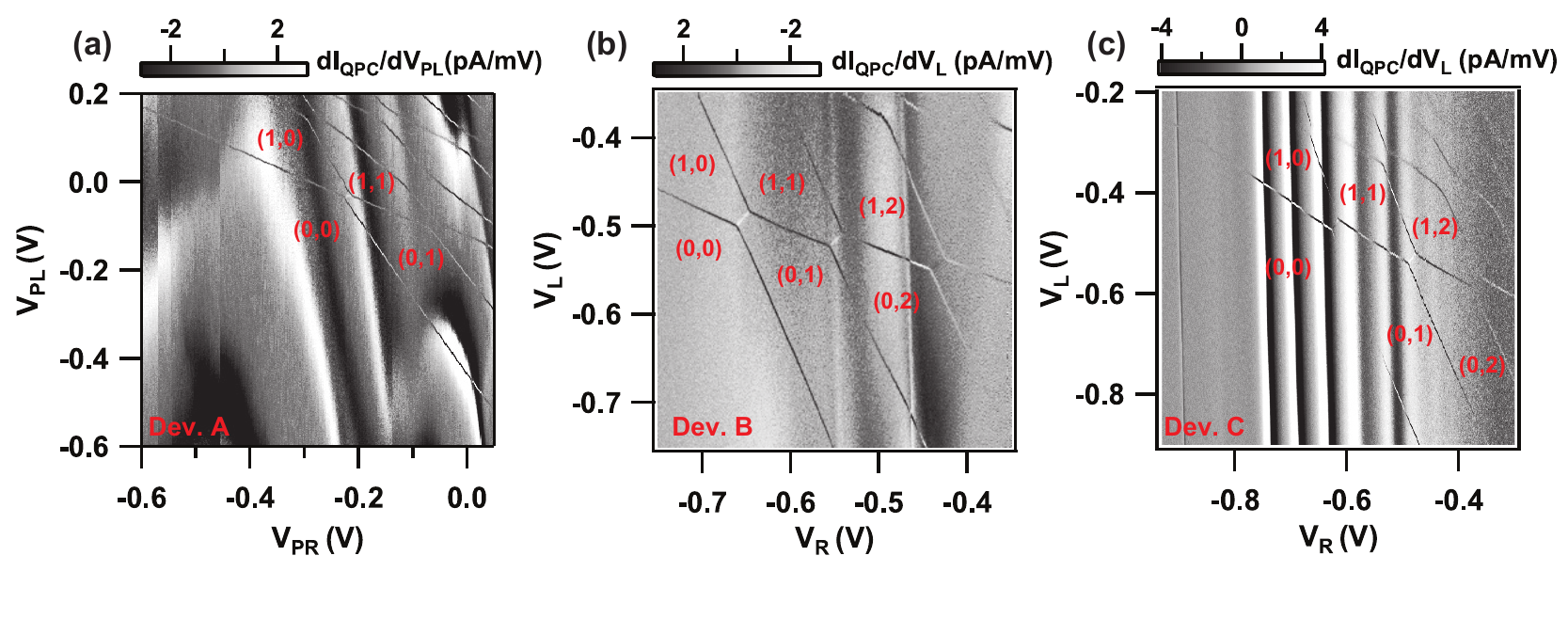}
\caption{\label{fig:Fig2} Charge stability diagrams are determined by measuring the transconductance of the QPC. Three different devices exhibit $(0,0)$ charge occupancy in the left and right dots respectively.  Each device has its gates biased accordingly: 
(a) Device A: 
$V_\text{FG}=2.25~\text{V}$, 
$V_\text{T}=0.33~\text{V}$, 
$V_\text{M}=-0.15~\text{V}$, 
$V_\text{L}=-0.325~\text{V}$, 
$V_\text{R}=-0.31~\text{V}$; 
(b) Device B: 
$V_\text{FG}=2.5~\text{V}$, 
$V_\text{T}=0.2~\text{V}$, 
$V_\text{M}=-0.3~\text{V}$, 
$V_\text{PL}=-0.35~\text{V}$, 
$V_\text{PR}=-0.35~\text{V}$;
(c) Device C: 
$V_\text{FG}=1.7~\text{V}$, 
$V_\text{T}=0.375~\text{V}$, 
$V_\text{M}=-0.1~\text{V}$, 
$V_\text{PL}=-0.075~\text{V}$, 
$V_\text{PR}=0.225~\text{V}$.}
\end{figure*}

In this letter we report low temperature measurements of three double quantum dot devices, fabricated on separate nominally undoped Si/SiGe heterostructure wafers. Measurements for each device demonstrate that the $(0,0)$ charge configuration is achieved. Pauli spin blockade has been observed in one of these devices at the $(1,1)\leftrightarrow(0,2)$ charge transition.

The device design is based on a double top-gated structure, similar to previous work,\cite{Xiao2010PRL} in which a global field gate is used to accumulate a two-dimensional electron gas (2DEG) in the Si well, and a set of localized depletion gates create a confining potential in which to form the quantum dots. The epitaxial structure, Fig.~\ref{fig:Fig1}(a), is grown by either chemical vapor deposition or molecular beam epitaxy on strain-relaxed $\text{Si}_{1-x}\text{Ge}_x$ ($x=0.33\pm0.02$) buffers. The field gate is isolated from the depletion gates via atomic layer deposition of dielectrics. Figure \ref{fig:Fig1}(b) shows a scanning electron micrograph of the electron-beam defined depletion gates with $\sim50~\text{nm}$ line widths. The depletion gates are designed to maximize the inter-gate distances, in particular the L-T and T-R tunnel barriers, while minimizing the area of the double quantum dot region. The addition of gate Q defines a quantum point contact (QPC) which is capacitively coupled to the dots and is able to sense the charge occupation of the device. 

Confirmation that a 2DEG is formed in the $10~\text{nm}$ Si well, and not at the Si-dielectric interface, was achieved via low-field Hall measurements ($B<0.5~\text{T}$) at $4.2~\text{K}$ using standard lock-in measurement techniques. Figures~\ref{fig:Fig1}(c) and \ref{fig:Fig1}(d) show Hall measurements of electron concentration and mobility from four Hall bars taken from the same processed wafer. A linear fit of $n_\text{2DEG}$ vs. $V_\text{FG}$, dashed line in Fig.~\ref{fig:Fig1}(c), yields a capacitance of $3.13\pm0.03\times10^{11}~\text{cm}^{-2}/\text{V}$, consistent with a 2DEG in the Si well; the computed capacitance for the Si well is $3.2\pm0.1\times10^{11}~\text{cm}^{-2}/\text{V}$ versus $3.8\pm0.1\times10^{11}~\text{cm}^{-2}/\text{V}$ for the Si-dielectric interface. Using the extrapolated threshold voltage, $0.36~\text{V}$, modeling indicates a relatively low quantity of fixed negative charge at the Si-dielectric interface, $\sim3\times10^{11}~\text{cm}^{-2}$. Figure~\ref{fig:Fig1}(d) shows the measured mobility, $\mu$, vs. $n_\text{2DEG}$, where $\mu\sim30,000~\text{cm}^2/\text{V-s}$ for $n_\text{2DEG}=5\times10^{11}~\text{cm}^{-2}$. The peak mobilities appear to be limited by $\sim10^{16}~\text{cm}^{-3}$ residual uniform background charge in our heterostructure consistent with theoretical mobility estimates\cite{Monroe1993} and a high saturation value of the Hall-inferred electron concentration, $n_\text{max}=6-8\times10^{11}~\text{cm}^{-2}$. Beyond $n_\text{max}$, electrons accumulate at the Si-dielectric interface.

Transport measurements were performed on three double quantum dot devices, each cooled down in a dilution refrigerator to a base temperature of $20~\text{mK}$. The field gate is biased such that devices are operated with $n_\text{2DEG}=4-6\times10^{11}~\text{cm}^{-2}$, a bias range that provides additional control over the tunnel barriers into the quantum dot. The depletion gates, T, L, PL, M, PR, R, are reverse biased in order to define a double dot. Tuning the device to zero charge occupancy is more easily achieved by measuring changes in the transconductance of the QPC corresponding to charge transitions in the double dot. Figure~\ref{fig:Fig2} shows AC differential transconductance measurements\cite{Croke2010} for all three devices plotted as a function of two depletion gate voltages. All exhibit the ability to achieve the $(0,0)$ charge configuration, that is zero charge occupancy in both left and right dots.\cite{Thalakulam2010} Both the $(0,0)\leftrightarrow(0,1)$ and $(0,0)\leftrightarrow(1,0)$ transition for each device show no indication of additional potential shifts associated with electron loading of the adjacent dot, for upwards of 4 addition energies. Table~\ref{tab:Devices} shows that all three devices have similar first addition energies $\Delta\mu_1=6-8~\text{meV}$ for both left and right dots along with similar gate lever arms, $\alpha$, for both L and R (PL and PR for device A). This consistency among the three devices, as well as among the individual dots within each double dot, strongly indicates that these dots are defined by the electrostatic potential created by the depletion gates and not by random disorder.

\begin{table}[h]
\begin{tabular}{|c|c|c|c|c|c|c|} % Column formatting, @{} suppresses leading/trailing space
%   \begin{tabular}{|p{0.5in}|p{0.4in}|p{0.4in}|p{0.4in}|p{0.4in}|p{0.4in}|p{0.4in}|}
\hline
     & \multicolumn{2}{|c|}{$\Delta\mu_1$ (meV)} & \multicolumn{2}{|c|}{$\alpha$ (eV/V)} & \multicolumn{2}{|c|}{$\Delta E_\text{S-T}$ (meV)} \\
%     & \multicolumn{2}{|c|}{energy (meV)} & \multicolumn{2}{|c|}{(eV/V)} & \multicolumn{2}{|c|}{} \\
\hline
  Device    & left & right & L(PL) & R(PR)  & (2,0)      & (0,2)      \\
\hline
    A       & 7.9  & 7.4  & 0.052  & 0.058  & $\sim$0.05 & $\sim$0.05 \\
\hline
    B       & 8.0  & 6.2  & 0.052  & 0.066  & $>$0.23    & $<$0.05    \\
\hline
    C       & 7.4  & 7.0  & 0.045  & 0.059  & ---        & 0.13       \\
\hline
Simulation  & 6.4  & 5.4  & 0.049  & 0.063  & \multicolumn{2}{c|}{$\leq$0.50} \\
\hline      
\end{tabular}
\caption{\label{tab:Devices} Measured and simulated parameters for the left and right dots of the double quantum dot. First addition energy, $\Delta\mu_1\equiv\mu_2-\mu_1$, for left and right dots. $\alpha$ is given for gates PL and PR for device A, and gates L and R for devices B and C. $\Delta E_\text{S-T}$ is the singlet-triplet energy splitting  of the $(2,0)$ or $(0,2)$ configurations (not measured for (2,0) of device C).}
\end{table}

\begin{figure}
\centering
\includegraphics[width=3.0in]{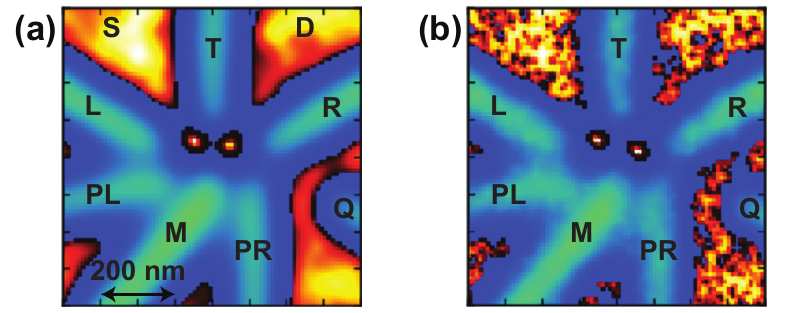}
\caption{\label{fig:Fig3} (a) Simulation of self-consistent electrostatic potential (blue) and electron density (red/yellow) of the nominal device tuned to the $(1,1)$ charge configuration, for an instance of discrete Si-dielectric interface charge. (b) Simulation of a similar device in which the source of the disorder potential is moved to within $10~\text{nm}$ of the quantum well.}
\end{figure}

The energetics of these devices is confirmed by real-space simulations using a hybrid Poisson-Schr\"odinger / full configuration interaction scheme. The total self-consistent charge density is partitioned into two regions, allowing us to solve the exact many-electron Hamiltonian for the few electron quantum dot system, while treating the bath regions semi-classically. Figure~\ref{fig:Fig3}(a) shows a simulation based on the nominal device design that includes $3\times10^{11}~\text{cm}^{-2}$ of fixed discrete charge, randomly distributed at the Si-dielectric interface, and $n_\text{2DEG}=4\times10^{11}~\text{cm}^{-2}$. The addition energies and $\alpha$ values for each of the two dots in this simulation are given in the last row of Table~\ref{tab:Devices}, showing consistency with the measured data. Figure~\ref{fig:Fig3}(b) shows a simulation of a similar device in which the source of the disorder potential is moved to within $10~\text{nm}$ of the electron layer, comparable to that typical of modulation doped Si/SiGe heterostructures. The increased influence of disorder both disturbs the quantum dot states and demonstrates the likelihood of disorder dot formation. 

The detection of spin blockade requires a singlet-triplet energy splitting, $\Delta E_\text{S-T}$, of the $(0,2)$ or $(2,0)$ state that is significantly larger than $kT$. Low-lying $N=2$ triplets in Si can be formed from either orbital excited states (for highly elongated configurations) or valley excited states. Magnetospectroscopy was performed on each of the dots within each device in order to infer $\Delta E_\text{S-T}$,\cite{Borselli2011} the results of which are given in Table~\ref{tab:Devices}. Two of the three devices, B and C, display the required energetics to enable a spin blockade measurement.

\begin{figure}[t]
\centering
\includegraphics[width=3.2in]{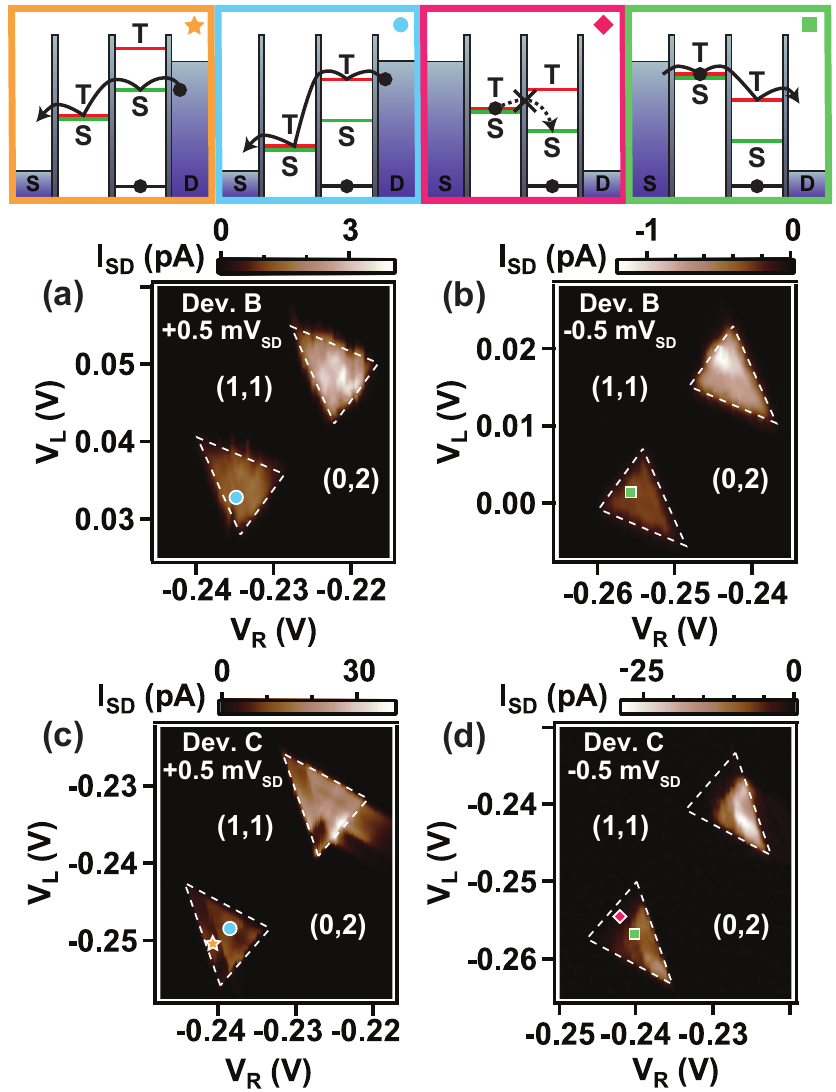}
\caption{\label{fig:Fig4} Through-dot current, $I_\text{SD}$ (pA), as a function of $V_\text{L}$ and $V_\text{R}$ for $(1,1)\leftrightarrow(0,2)$ charge transitions for devices B and C with schematic energy diagrams.  (a) and (b) show no observable Pauli spin blockade for device B as expected from the small $\text{S}(0,2)-\text{T}(0,2)$ energy splitting. (c) Device C at $V_\text{SD}=+0.5~\text{mV}$; current is observed throughout both bias triangles (white dashed lines).  The orange star indicates the region in which only $\text{S}(0,2)\rightarrow\text{S}(1,1)$ transitions are allowed.  The blue circle represents the region of increased current once the $\text{T}(0,2)$ state becomes energetically accessible.  (d) Device C at $V_\text{SD}=-0.5~\text{mV}$, current is suppressed in the region marked with a red diamond due to Pauli spin blockade until the $\text{T}(1,1)\rightarrow\text{T}(0,2)$ transition becomes allowed (green square).}
\end{figure}

The gate design provides the ability to tune the tunnel barriers with minimal effect on the confining gate potentials, allowing us to observe transport in bias triangles\cite{Hanson2007} down to the $(0,2)$ charge configuration, similar to what has been previously achieved in GaAs.\cite{Johnson2005} The asymmetric charge sensor design limited our ability to observe transport at the $(2,0)\leftrightarrow(1,1)$ charge transitions, preventing us from attempting a blockade measurement at that transition in device B. Figure~\ref{fig:Fig4} presents transport data for both forward and reverse DC biases at the $(1,1)\leftrightarrow(0,2)$ charge transition for devices B and C. From the magnitudes of the measured currents, we infer that the limiting tunnel times are $<100~\text{ns}$ (Dev. B) and $<10~\text{ns}$ (Dev. C). Figures~\ref{fig:Fig4}(a) and \ref{fig:Fig4}(b) show bias triangles for device B where blockade is not seen, as expected due to the small $\Delta E_\text{S-T}$ of its $(0,2)$ charge configuration; similarly no blockade was observed in device A. The signature of Pauli spin blockade in device C can clearly be seen by comparing the size of the transport region in the bias triangles in Figs.~\ref{fig:Fig4}(c) and \ref{fig:Fig4}(d) with $\pm0.5~\text{mV}$ applied across the dots. The size of the blockade region in the reverse bias configuration implies the presence of an excited $(0,2)$ triplet with a splitting of $0.14\pm0.01~\text{meV}$ from the ground state, consistent with the magnetospectroscopy-measured $\Delta E_\text{S-T}=0.13\pm0.01~\text{meV}$ for this device. 

In summary we have fabricated and measured double quantum dots that are easily tuned to the last electron.  Isolating the electron system from sources of charged impurities, such as those at an oxide interface or a doping layer, has resulted in devices with greatly improved stability and reproducible performance. Pauli spin blockade at the $(1,1)\leftrightarrow(0,2)$ charge transition has been observed with $<10~\text{ns}$ tunneling times to the electron baths, enabling the future demonstration of coherent control of the two-electron spin system in Si.

%\begin{acknowledgments}
The authors gratefully acknowledge Profs. H.~W.~Jiang and C.~Anderson for useful discussions. 
Sponsored by United States Department of Defense. The views and conclusions contained in this document are those of the authors and should not be interpreted as representing the official policies, either expressly or implied, of the United States Department of Defense or the U.S. Government. 
Approved for public release, distribution unlimited.
%\end{acknowledgments}

\end{document}